\documentclass[a4paper,12pt]{article}
\pdfoutput=1
\usepackage{graphicx}
\usepackage{epstopdf}
\usepackage{amssymb,amsmath} 
\begin{document}
\title{Continuation and stability deduction\\ of resonant periodic orbits\\ in three dimensional systems\footnote{Proceedings of the $6^{th}$ International Conference on Numerical Analysis (NumAn $2014$). Published by the Applied Mathematics and Computers Lab, Technical University of Crete (AMCL/TUC), Greece.}}

\author{Kyriaki I. Antoniadou, George Voyatzis and Harry Varvoglis\\
\small{Section of Astrophysics, Astronomy and Mechanics}, \\\small{Department of Physics, Aristotle University of Thessaloniki,} \\
\small{Thessaloniki, 54124, Greece}\\
\small{kyant@auth.gr, voyatzis@auth.gr and varvogli@physics.auth.gr}}

\maketitle
\begin{abstract}
\noindent
In dynamical systems of few degrees of freedom, periodic solutions consist the backbone of the phase space and the determination and computation of their stability is crucial for understanding the global dynamics. In this paper we study the classical three body problem in three dimensions and use its dynamics to assess the long-term evolution of extrasolar systems. We compute periodic orbits, which correspond to exact resonant motion, and determine their linear stability.  By computing maps of dynamical stability we show that stable periodic orbits are surrounded in phase space with regular motion even in systems with more than two degrees of freedom,  while chaos is apparent close to unstable ones. Therefore, families of stable periodic orbits, indeed, consist backbones of the stability domains in phase space.
\end{abstract}

\section{Introduction}
Many extrasolar multiplanet systems discovered in the past 20 years seem to move on Keplerian ellipses with periods that satisfy rational relations. Such a type of resonant motion is a consequence of the existence of periodic orbits, and, particularly, stable periodic orbits of the same resonance. Many studies of resonances in the planar three body problem (TBP) can be found in literature (see e.g. \cite{Micht06,Hadjidem06}). Recently periodic orbits for the spatial TBP have been also presented \cite{AV14}.  

In Hamiltonian systems, it is known that in phase space the stable periodic orbits are surrounded by invariant tori, while in the neighborhood of unstable periodic orbits homoclinic webs are formed that give rise to chaotic motion. Applying this notion to planetary dynamics, we may claim that chaos destabilizes planetary systems and therefore a real planetary system can be found for initial conditions only in phase space regions of regular motion \cite{voyatzis08}.  Although regular obits are not necessarily associated with stable periodic orbits, the last ones are very important in planetary dynamics because (i) many systems consist of planets that move on highly eccentric orbits, which can survive collisions only when the system evolves close to an exact resonance i.e. close to a periodic orbit and (ii) stable periodic orbits become attractors for a planetary system that moves  inside the protoplanetary disk of its early stage of formation \cite{Lee02}.

In this study we show how we can determine orbits of long-term stability in the spatial three body problem.  After having computed families of periodic orbits and determined their linear stability, we construct maps of dynamical stability by using a chaotic indicator. In this way we identify the extent of stable regions in phase space.

\section{Model and periodic orbits}
We consider the  general three body problem, consisting of a star, $S$, with mass $m_0$, and two planets, $P_1$ and $P_2$ with masses $m_i\ll m_0$, $i=1,2$, normalized such that $\sum_{i=0}^2 m_i=1$. In the next, indices 0, 1 and 2 always indicate quantities of the star, the inner and the outer planet, respectively. By introducing a rotating frame of reference $Gxyz$, which rotates around the constant angular momentum vector and contains always the bodies $S$ and $P_1$ in the plane $Gxz$, the position of the system is determined  by the four variables $(x_1, x_2, y_2, z_2)$ (see  \cite{mich79,AV14}). The differential equations of motion are 
\begin{equation}
\begin{array}{l}
\ddot x_1=-\frac{m_0 m_2 (x_1-x_2)}{\mu r_{12}^3}-\frac{m_0 m_2 ( a x_1+x_2)}{\mu r_{02}^3}-\frac{\mu x_1}{r_{01}^3}+x_1\dot{\theta}^2\\[0.3cm] 
\ddot x_2=\frac{m_1 (x_1-x_2)}{\mu r_{12}^3}-\frac{m_0 (a x_1+x_2)}{\mu r_{02}^3}+x_2\dot{\theta}^2+2 \dot y_2 \dot{\theta}+y_2\ddot{\theta}\\[0.3cm]
\ddot y_2=-\frac{m_1 y_2}{\mu r_{12}^3}-\frac{m_0 y_2}{\mu r_{02}^3}+y_2\dot{\theta}^2-2 \dot x_2 \dot{\theta}-x_2\ddot{\theta}\\[0.3cm]
\ddot z_2=\frac{m_1 (z_1-z_2)}{\mu r_{12}^3}-\frac{m_0 (a z_1+z_2)}{\mu r_{02}^3}
\label{seq}
\end{array}
\end{equation}
where $a=m_1/m_0$, $b=m_2/m$, $\mu=m_0 + m_1$ and $r_{ij}$ is the distance between the bodies indicated by the indices $i$ and $j$ and is a function of the system's variables. The quantities $\dot \theta$ and $\ddot \theta$ can be expressed also as functions of the system's variables and the constant angular momentum.  

For the system (\ref{seq}) we can define periodic solutions of period $T$, $\mathbf{Q}(T)=\mathbf{Q}(0)$, where $\mathbf Q$ refers to the variables of position and velocity. By studying the evolution on a Poincar\'e map, defined e.g. by the surface of section $y_2=0$ with $\dot y_2>0$, the set of initial conditions of periodic orbits 
$$
\mathbf{Q(\rm{0})}=\{x_{10},x_{20},y_{20}=0, z_{20},\dot x_{10},\dot x_{20}, \dot y_{20},\dot z_{20}\} 
$$
forms characteristic curves (or families of periodic orbits) in the phase space of the Poincar\'e map, due to mono-parametric continuation. 
The linear stability of periodic orbits is derived from the variational equations and their solution (see e.g. \cite{skokos01})
\begin{equation}
\begin{array}{ccc}
\dot{\vec{\xi}}=\textbf{J}(t)\vec{\xi} & \Rightarrow & \vec{\xi}=\Delta(t)\vec{\xi}_0,
\end{array}
\end{equation}
where \textbf{J} is the Jacobian of the right part of system's equations and $\Delta (t)$ the fundamental matrix of solutions.

According to Floquet's theory, the deviations $\vec{\xi}(t)$ remain bounded iff all the eigenvalues of $\Delta (T)$ (which form conjugate pairs) are distinct and lie on the complex unit circle. Then, the corresponding periodic orbit is {\em linearly stable}. If at least one pair of conjugate eigenvalues lies outside of the complex unit circle the periodic orbit is {\em linearly unstable}. We remark that due to the energy integral, it is always $\lambda_1=\lambda_2=1$ but this pair is not taken into account.

Planar periodic orbits exist in the context of the planar model ($z_2=\dot z_2=0$). In this case the monodromy matrix of stability can be divided in two components, one indicating the {\em horizontal stability} and the other the {\em vertical stability} \cite{mich79}. When the eigenvalues that refer to the component of vertical stability are equal to 1, the corresponding planar periodic orbit is called {\em vertically critical orbit} ({\em vco}). Families of three dimensional periodic orbits bifurcate at {\em vco}. 

\begin{figure}
\begin{center}
\includegraphics[width=12cm]{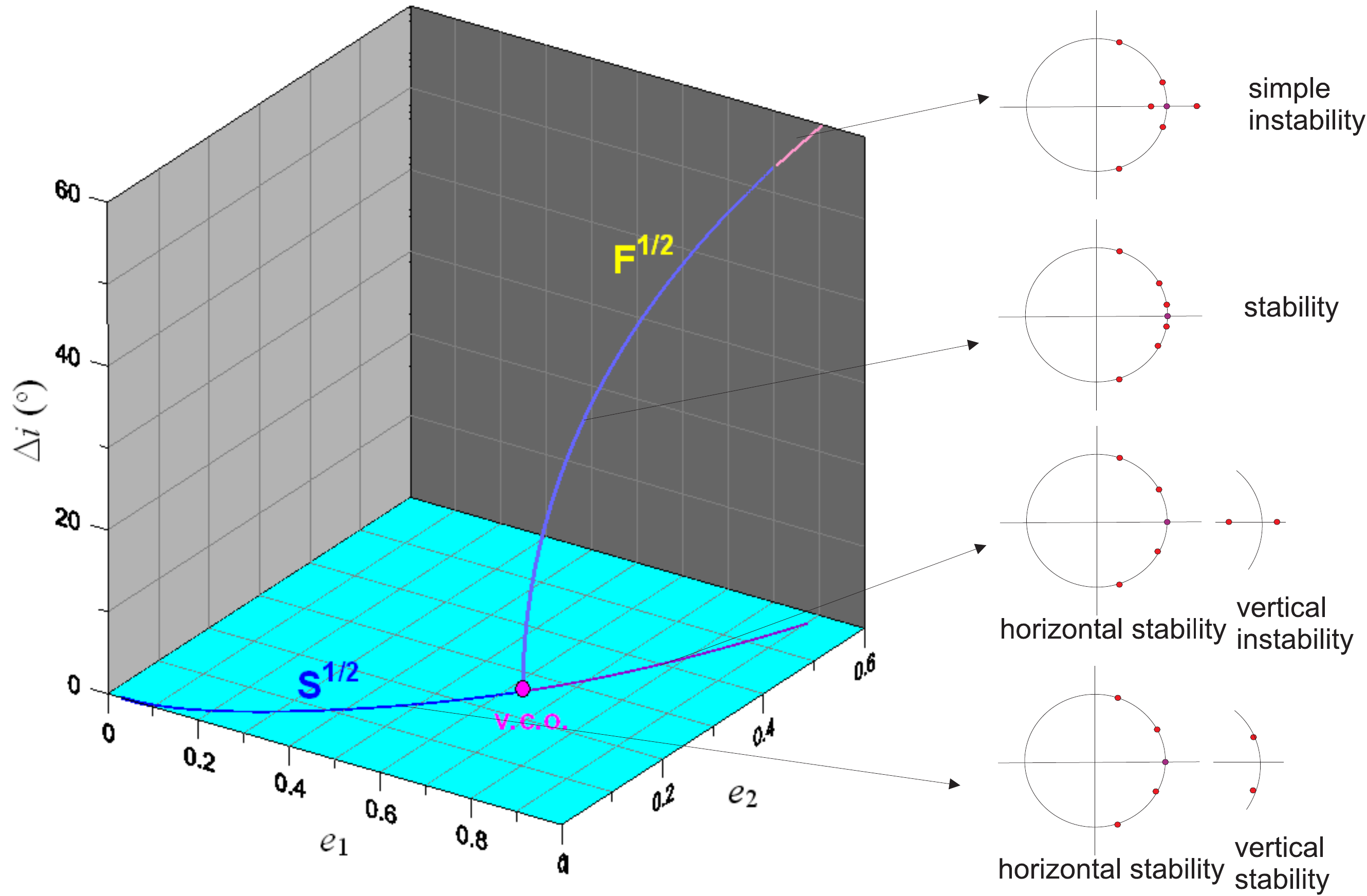}
\end{center}
\caption{Families of 2/1 resonant periodic orbits for $m_1=0.001$, $m2=0.002$. $S^{1/2}$ is the family of planar orbits and $F^{1/2}$ the family of 3-dimensional orbits. The qualitative distribution of the eigenvalues on the complex unit circle is given for the various parts of families with different stability type.}
\label{figFams}
\end{figure}

In Fig. \ref{figFams} we present a typical example of the above mentioned notions. We describe the orbits by considering the orbital elements of the two planets, namely the semimajor axes $a_i$, the eccentricities $e_i$, the inclinations $i_i$, the arguments of pericenter $\omega_i$, the longitudes of nodes $\Omega_i$ and the mean anomalies $M_i$. For all cases presented in the following it is $\omega_1=90^\circ$, $\omega_2=270^\circ$, $\Omega_1=270^\circ$, $\Omega_2=90^\circ$ and $M_1=0^\circ$, $M_2=180^\circ$.

For $m_1=0.001$ and $m_2=0.002$  and 2/1 resonance ($a_1=1$ and $a_2\approx1.58$) we computed for the planar model the family of periodic orbits $S^{1/2}$ \cite{Hadjidem06}, which is projected on the plane of eccentricities of Fig. \ref{figFams}. The family is horizontally stable. It starts also as vertically stable but at $e_1=0.63$, $e_2=0.256$ becomes vertically unstable. From this point we obtain the bifurcation of the 3-dimensional family $F^{1/2}$, which is presented in the space $e_1 - e_2 -\Delta i$, where $\Delta i=i_2+i_1$ is the mutual inclination between the planets. 
This family starts as stable but for high mutual inclinations ($\Delta i>47^\circ$) becomes unstable. In the same figure we show, in a qualitative sense, the distribution of the eigenvalues of the monodromy matrix (red dots) which correspond to parts of the families with different stability type.                 

\section{Chaos indicator and maps of dynamical\\ stability}
The Fast Lyapunov Indicator (FLI) \cite{froesle97} is one of the most simple indicators that can classify an orbit as regular or chaotic. Also it has been proved to be very efficient for the particular model of the TBP \cite{voyatzis08}. It is defined in various ways. Here we use it in a simple form, called also {\em de-trended} FLI, 
$$
DFLI(t)=log \left ( \frac{1}{t}|\mathbf{\xi}(t)| \right ),
$$
where $\mathbf{\xi}$ is a deviation vector derived from the integration of variational equations along the orbit. For regular orbits 
{\em DFLI} tends to a constant value as $t\rightarrow \infty$. In case of a periodic orbit this value is small (of order 1). For chaotic orbits it increases exponentially. In the TBP, our computations have shown that performing numerical integrations for $10^5$ - $2\,10^5$ t.u.,  regular orbits are characterized by a final {\em DFLI} value less than 10. On the other hand, for chaotic orbits we found that if {\em DFLI} takes a value larger than 15 then it continues to increase steeply. We stop the integration before the time limit mentioned above, when $DFLI>30$. 

\begin{figure}
\begin{center}
$\begin{array}{@{\hspace{-.2em}}cc}
\includegraphics[width=6cm]{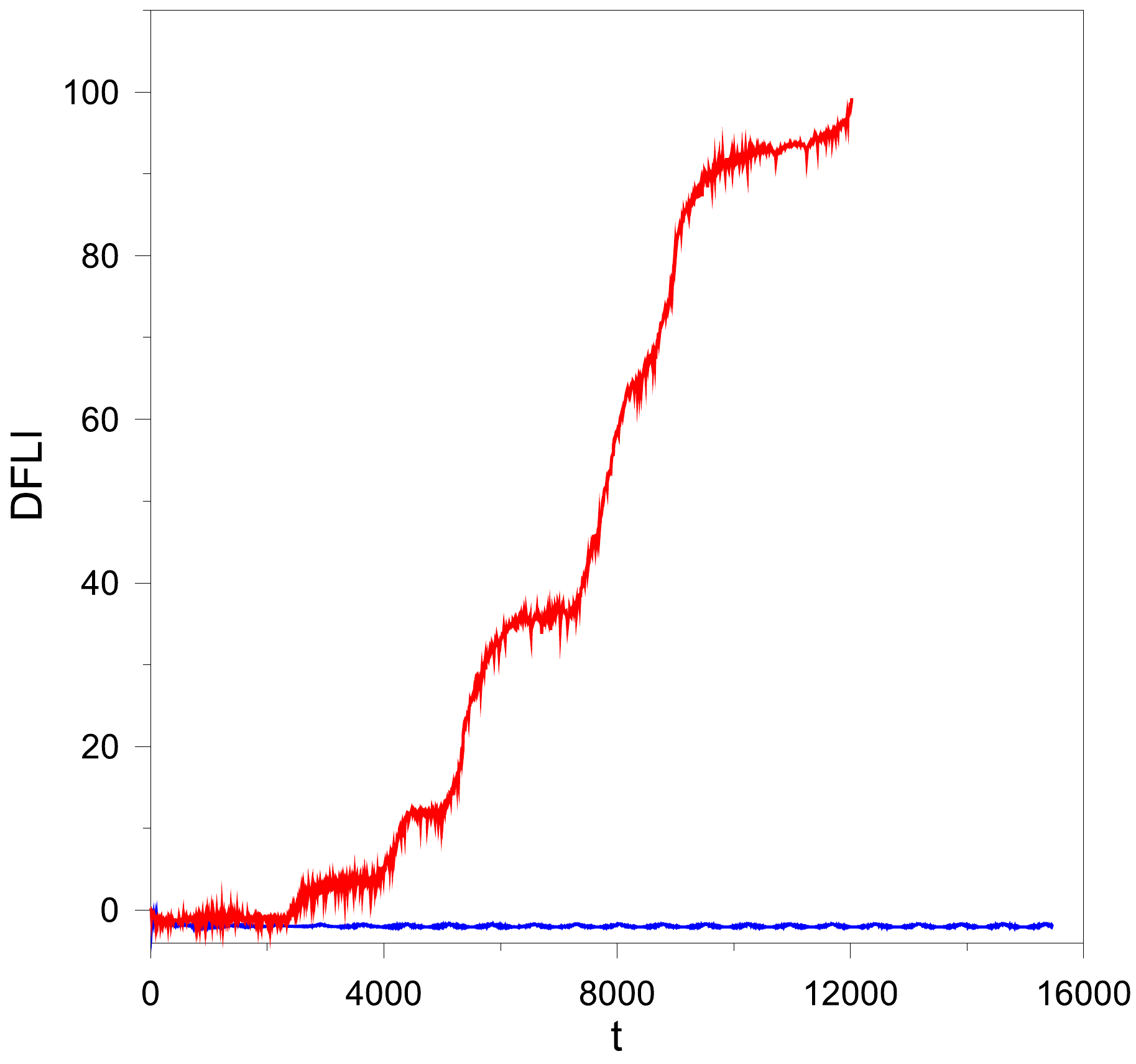}  &\includegraphics[width=6cm]{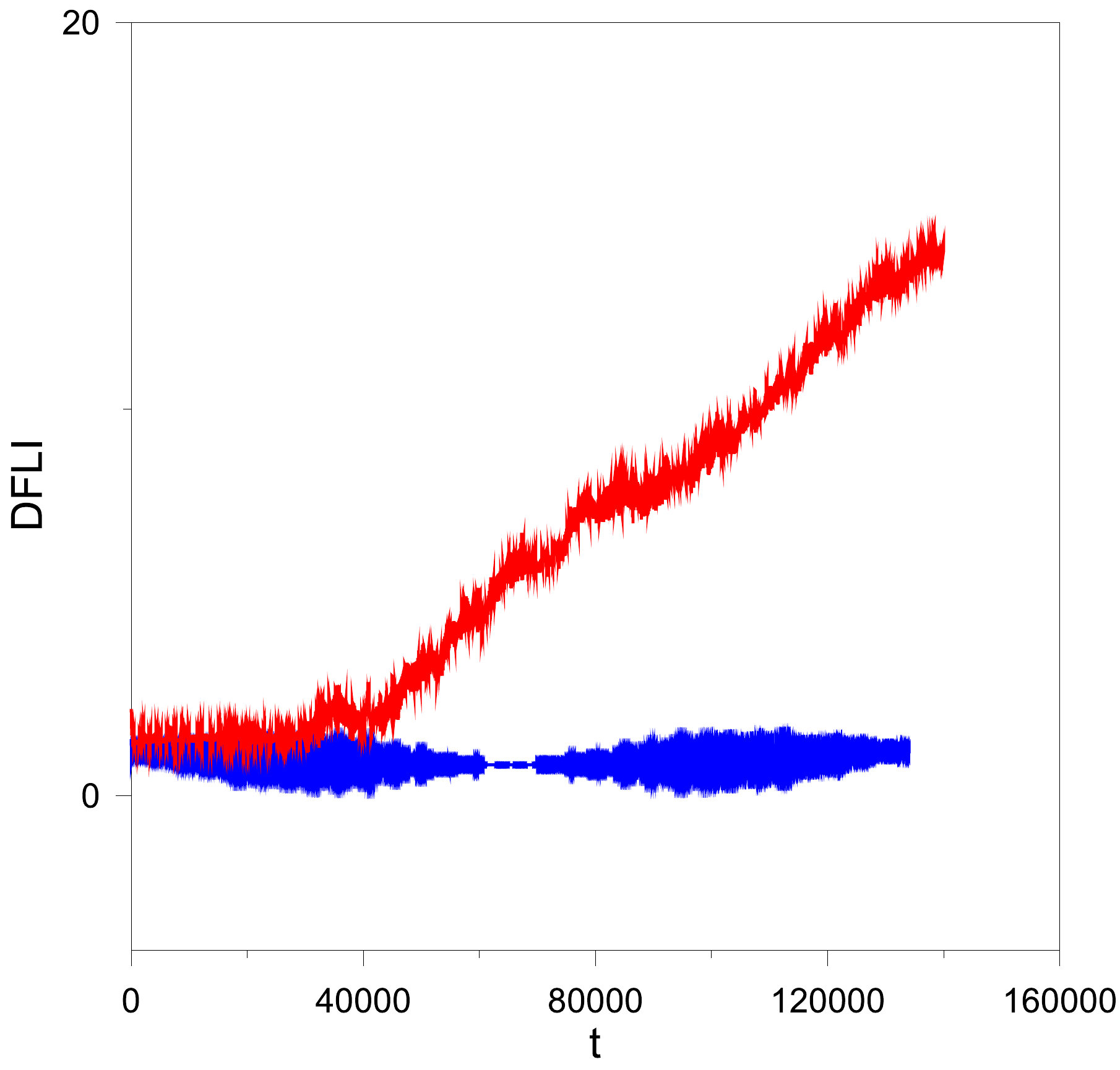}\\
\textnormal{(a)}  &\textnormal{(b)}\\
\end{array} $
\end{center}
\caption{The evolution of DFLI. Blue coloured lines correspond to the regular evolution, while the red ones to chaotic evolution.}
\label{figDFLI}
\end{figure}
    
\begin{figure}[tb]
\begin{center}
$\begin{array}{@{\hspace{-.2em}}ccc}
\includegraphics[width=5.9cm]{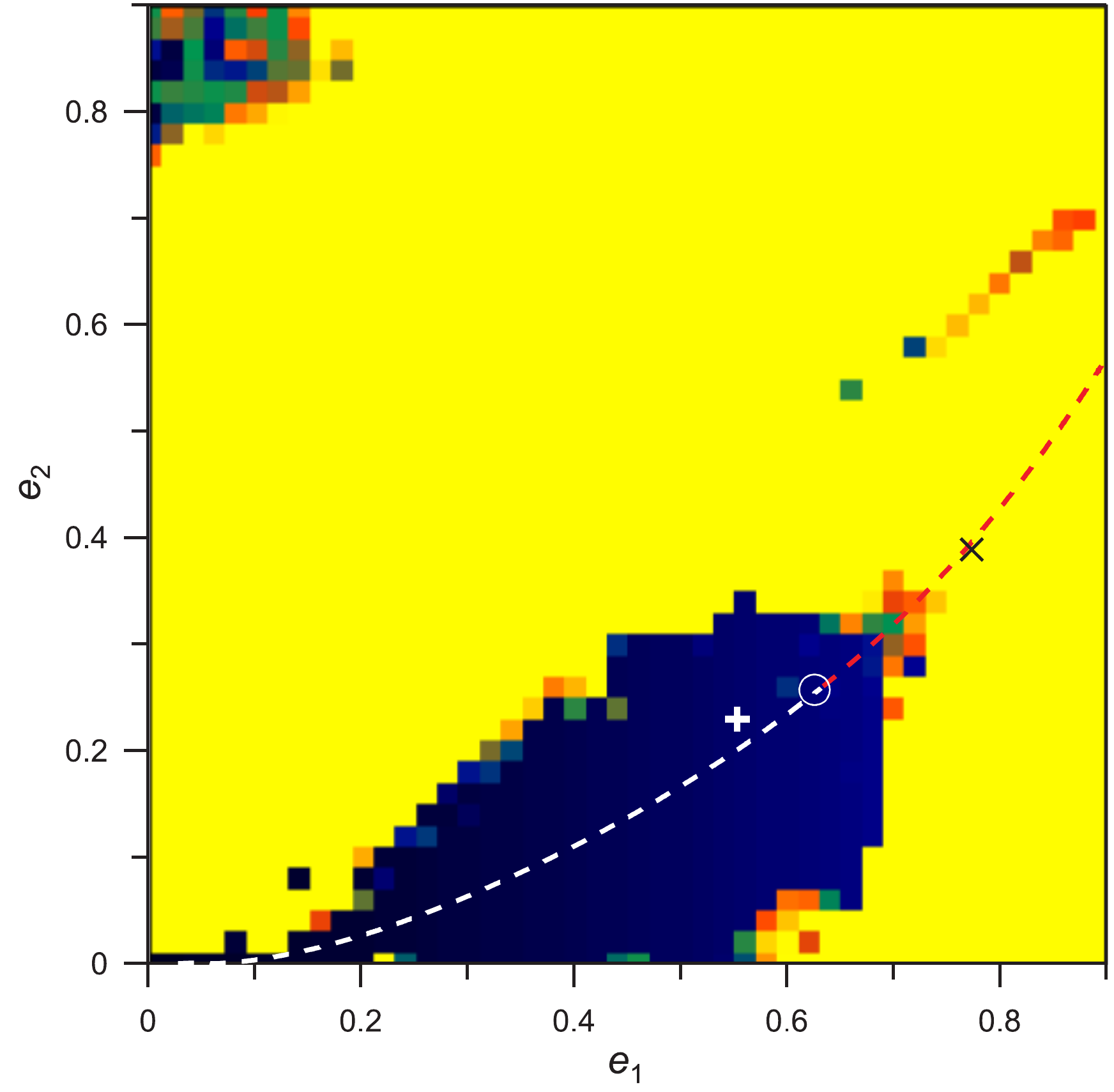} &
\includegraphics[height=3cm]{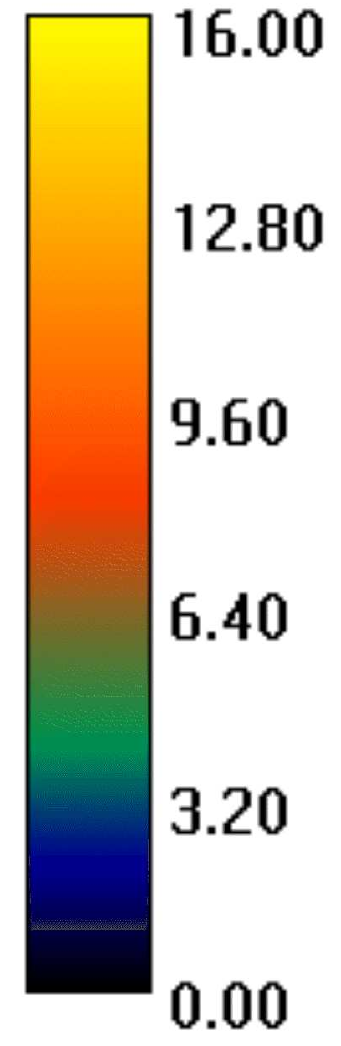} &\includegraphics[width=5.9cm]{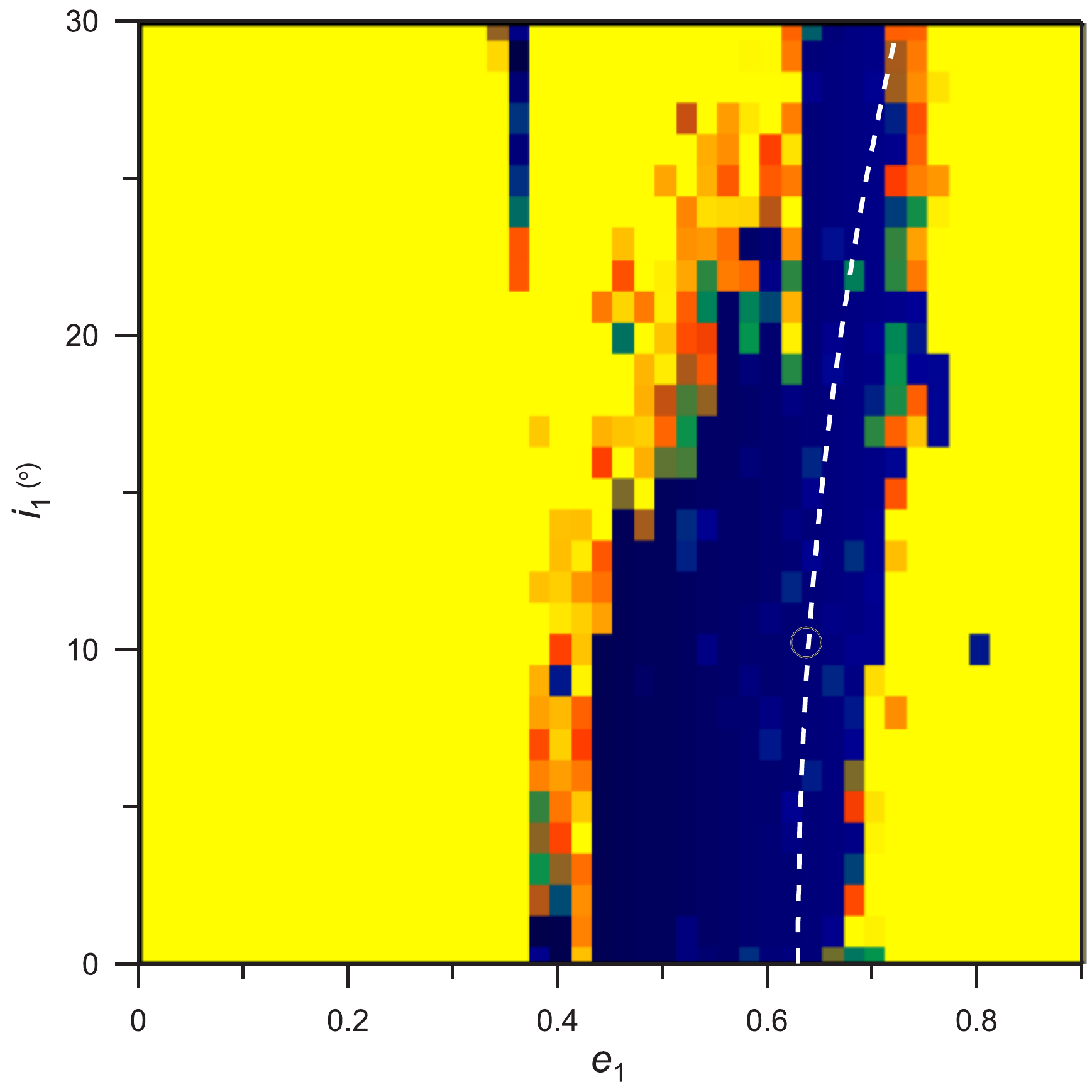} \\
\textnormal{(a)} &  &\textnormal{(b)}\\
\end{array} $
\end{center}
\caption{Maps of dynamical stability in the planes {\bf a} $(e_1,e_2)$ and {\bf b} $(e_1,i_1)$. The families of periodic orbits are also presented.}
\label{figMaps}
\end{figure}

In Fig. \ref{figDFLI} we show the evolution of {\em DFLI} for a regular and a chaotic orbit. In panel (a) the regular orbit has initial conditions very close to the  stable periodic orbit (see Fig. \ref{figFams}) with $e_1=0.630$, $e_2=0.257$ and $Di=4^\circ$. The chaotic orbit starts with initial conditions very close to the unstable periodic orbit with $e_1=0.825$, $e_2=0.486$ and $Di=52^\circ$. In panel (b)  the evolution of {\em DFLI} for a regular and a chaotic orbit is shown as well. Their initial conditions are indicated on the dynamical map of Fig.\ref{figMaps}a with ($+$) and ($\times$), respectively. 
    
Considering a grid of initial conditions on a plane defined by two variables of the system, and fixing the other variables at values that correspond to those of a periodic orbit, we compute, for $t=t_{max}$, the {\em DFLI} of each orbit in the grid. So, in the 8-dimensional phase space of the system, we obtain particular 2-dimensional sections, named {\em maps of dynamical stability}, that depict the distribution of regular and chaotic orbits. Two representative maps, which were computed for $t_{max}=150000$ t.u., are given in Fig. \ref{figMaps}. Dark or light colors represent regular or chaotic orbits, respectively.

In panel (a) of Fig. \ref{figMaps} we present a dynamical map on the eccentricity plane, which corresponds to a mutual planetary inclination of $10^\circ$. The family of planar periodic orbits is also presented. The family is horizontally and vertically stable up to the point indicated by the `circle' symbol ({\em vco}). Up to this point, the family is surrounded by regular orbits and the evolution of the system shows long-term stability. Beyond the {\em vco}, although the planar family is horizontally stable, it becomes vertically unstable. Such vertical instability causes the orbits to become irregular and the stability region terminates. 

In panel (b) of Fig. \ref{figMaps} we compute a dynamical map on the plane $e_1\,-\, i_1$. The family of three-dimensional periodic orbits is also projected in this space. We note that the initial conditions of  all orbits of the family cannot be depicted exactly on this plane because $e_2$ and $i_2$ vary slightly along the family. We clearly observe that the family of stable periodic orbits is actually the backbone of the region with regular orbits.

\section{Conclusions}
In the paper we discussed the existence and the bifurcation of periodic orbits from the planar to the 3-dimensional TBP. The type of their linear stability can be determined from the distribution of the eigenvalues of the monodromy matrix on the complex unit circle. 

Even for high dimensional systems, the existence of stable periodic orbits in phase space indicates the existence of stability regions too, i.e. domains of initial conditions for which the system evolves regularly for very long time spans. So the computation of periodic orbits can be used as a method to determine regular motion. By using the chaos indicator {\em DFLI}, we computed maps of dynamical stability and showed that families of stable periodic orbits define backbones of stability domains in phase space.    

The present model can be applied to observed exoplanetary systems since by using the presented analysis we determine possible configurations for these systems. 

\vspace{0.5cm}
{\bf Acknowledgments} This research has been co-financed by the European Union (European Social Fund - ESF) and Greek national funds through the Operational Program ``Education and Lifelong Learning'' of the National Strategic Reference Framework (NSRF) - Research Funding Program: Thales. Investing in knowledge society through the European Social Fund.


\begin{thebibliography}{99} 

\bibitem{Micht06}
Michtchenko, T.A., Beaug\'e, C., Ferraz-Mello, S.: Stationary orbits in resonant extrasolar planetary systems.
Celest. Mech. Dyn. Astron. 94, 411-–432 (2006)

\bibitem{Hadjidem06}
Hadjidemetriou, J.D.: Symmetric and asymmetric librations in extrasolar planetary systems: a global view.
Celest. Mech. Dyn. Astron. 95, 225-–244 (2006)

\bibitem{AV14}
Antoniadou, K.I., Voyatzis, G, : Resonant periodic orbits in the exoplanetary systems, Astrophys. Sp. Sci., 349, pp.657--676 (2014).

\bibitem{voyatzis08}
Voyatzis G.: Chaos, Order and periodic orbits in the 3:1 resonant planetary dynamics, ApJ, 675, 802--816 (2008)

\bibitem{Lee02}
Lee, M.H., Peale, S.J.: Dynamics and origin of the 2:1 orbital resonances of the GJ 876 planets. ApJ, 567, 596-–609 (2002)

\bibitem{mich79}
Michalodimitrakis,M.: On the continuation of periodic orbits from the planar to the three-
dimensional general three-body problem. Celest. Mech. 19, 263--277 (1979)

\bibitem{skokos01}
Skokos, C. : On the stability of periodic orbits of high dimensional autonomous Hamiltonian systems
Physica D, 159, No 3-4, 155-179 (2001)

\bibitem{froesle97}
Froeschl\'e, C., Lega, E., Gonczi, R., Fast Lyapunov Indicators. Application to Asteroidal Motion, Celest. Mech. Dyn. Astron., 67, 41--62 (1997)

\end{thebibliography}
\end{document}